\documentclass[conference]{IEEEtran}
\IEEEoverridecommandlockouts
\usepackage{cite}
\usepackage{amsmath,amssymb,amsfonts}
\usepackage{graphicx}
\usepackage{textcomp}
\usepackage{xcolor}
\def\BibTeX{{\rm B\kern-.05em{\sc i\kern-.025em b}\kern-.08em
    T\kern-.1667em\lower.7ex\hbox{E}\kern-.125emX}}
\begin{document}

\title{Machine learning architectures to predict motion sickness using a Virtual Reality rollercoaster simulation tool}

\author{\IEEEauthorblockN{Stefan Hell}
\IEEEauthorblockA{\textit{Kingston University} \\
London, \\United Kingdom \\
k1729468@kingston.ac.uk}
\and
\IEEEauthorblockN{Vasileios Argyriou}
\IEEEauthorblockA{\textit{Kingston University} \\
London, \\United Kingdom \\
Vasileios.Argyriou@kingston.ac.uk}
}
\maketitle

\begin{abstract}
Virtual Reality (VR) can cause an unprecedented immersion and feeling of presence yet a lot of users experience motion sickness when moving through a virtual environment. Rollercoaster rides are popular in Virtual Reality but have to be well designed to limit the amount of nausea the user may feel. This paper describes a novel framework to get automated ratings on motion sickness using Neural Networks. An application that lets users create rollercoasters directly in VR, share them with other users and ride and rate them is used to gather real-time data related to the in-game behaviour of the player, the track itself and users' ratings based on a Simulator Sickness Questionnaire (SSQ) integrated into the application. Machine learning architectures based on deep neural networks are trained using this data aiming to predict motion sickness levels. While this paper focuses on rollercoasters this framework could help to rate any VR application on motion sickness and intensity that involves camera movement. A new well defined dataset is provided in this paper and the performance of the proposed architectures are evaluated in a comparative study.
\end{abstract}

\begin{IEEEkeywords}
VR, Rollercoasters, Motion Sickness, Neural Network
\end{IEEEkeywords}

\section{Introduction}

\begin{figure}[h!]
		\centering
  \includegraphics[scale=0.3]{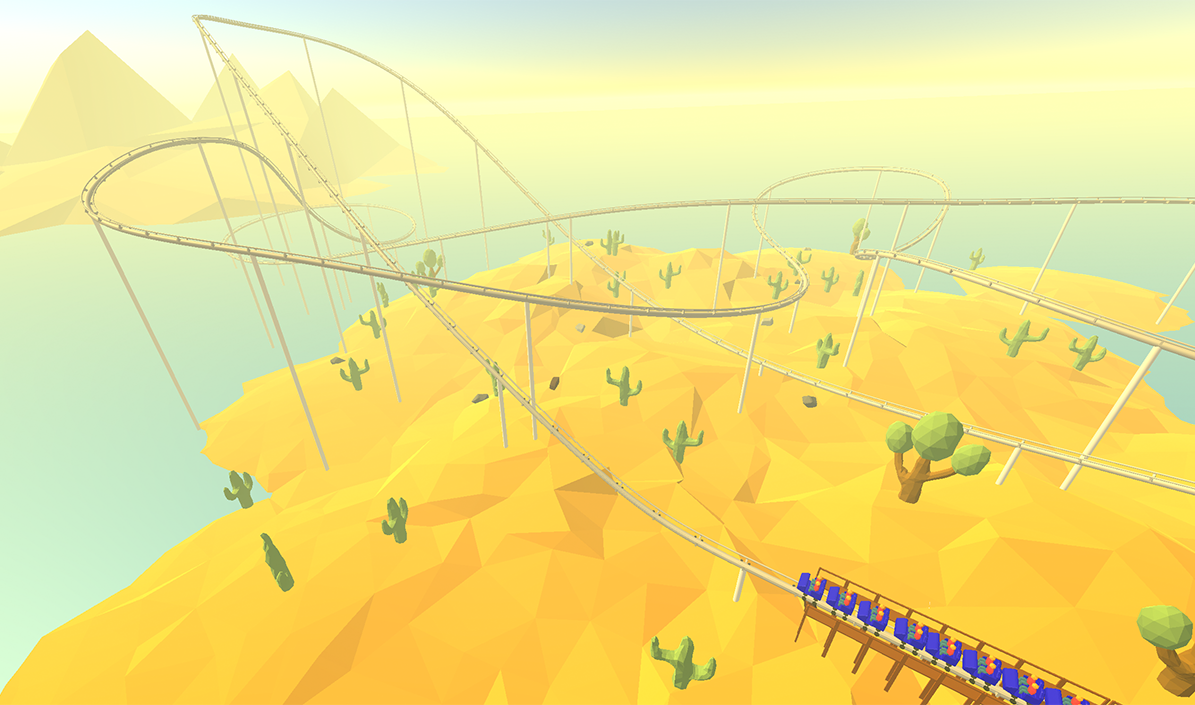}
  \caption{Example of a rollercoaster designed in the VR application used to gather data.}
  \label{fig:rollercoaster}
\end{figure}

Due to advances in technology and hardware Virtual Reality (VR) has gained popularity in the last years. Virtual Reality has also been utilized in domains outside video games including health care, human computer interaction, behaviour and action analysis, construction and architecture \cite{b7,b12,b6}. VR displays and devices fully immerse the user in the virtual environment by blocking other visual inputs that might disturb the experience. Unfortunately, this level of presence also leads to the undesirable side effect of motion sickness which is a set of unpleasant symptoms due to the exposure to a virtual environment and can last from few minutes to even days. These symptoms include eyestrain, headache, nausea or even vomiting \cite{b8}.

This paper studies the effect of in-game behaviour on motion sickness. The research problem is described as: can we predict if a given unknown virtual environment and a user's in-game behaviour may create motion sickness using supervised learning methods.

Rollercoaster rides are very popular in Virtual Reality (VR) as the feeling of presence leads to a much more realistic experience than watching an on-ride video on a monitor. In VR the sensation of acceleration, speed and height is much more realistic. A novel framework to facilitate the creation and motion sickness evaluation of new rollercoasters using VR is introduced in this paper – similar to a flight simulator you can fly through the air by tilting your head (see Figure \ref{fig:steering}).

Well-designed rollercoasters exert gravitational forces (g-forces) at the right intensity on the user to provide an intense ride while at the same time causing as little motion sickness as possible. Classic theme park simulations like Rollercoaster Tycoon use sophisticated algorithms to determine excitement, intensity and nausea using manually defined parameter thresholds and ad-hoc ranges. This paper describes a novel approach to automatically rate rollercoasters with neural networks. The new RCVR database of rollercoasters and ratings is introduced and used to train the proposed neural network architectures to rate any new rollercoaster created by a user mainly in terms of motion sickness.

While rollercoasters are a rather specific use case the proposed framework could be used for a wide range of VR applications and games in the future. Any VR application that involves movement through an environment could be evaluated on the amount of motion sickness it induces on the average user. This is especially important as developers and testers who use VR on a daily basis become resistent to motion sickness and while they feel fine testing it the average users may not. The proposed framework could provide reliable and low cost estimates. In summary, our contributions are:
\begin{itemize}
\item A novel framework for motion sickness prediction using in-game behaviour and environmental data.
\item A new dataset with real time data series and ground truth labels based on an integrated questionnaire.
\item A set of different input time series and values were considered in this work including raw information and meta-data.
\item We achieve this via a simple architecture that includes FC, recurrent and pooling layers.
\item We report results for a number of experiments on the proposed RCVR dataset, illustrating that our framework provides reasonable and promising results for motion sickness prediction.
\end{itemize}
At section 2 previous work relevant to VR and motion sickness is presented. At section 3, the proposed framework is presented and Section 4 discusses the findings of this work. Section 6 concludes the study.

\section{Literature review}

Motion sickness is a major barrier on the way to mass success of Virtual Reality and is therefore a leading topic in research. Physically induced motion sickness can appear in a wide range of occasions, for example when you get seasick on a boat. In Virtual Reality users experience a similar form of motion sickness called visually induced motion sickness (VISM) that results in similar symptoms. They experience physically induced motion sickness as well as blurred vision and headaches \cite{b3}. As Virtual Reality can create a much stronger sensation of presence compared to a monitor – the users fees like they are actually in the virtual world – the brain expects a force exerted on their body on any acceleration or deceleration. As these forces are missing there is a discrepancy between the visual and physical experience can lead to motion sickness. Tackling this problem is one of the most important challenges to ensure a really immersive experience in Virtual Reality.

Some studies limit motion sickness by including arm or foot movement \cite{b4} while others focus on limiting the field of view during movement or providing a frame of reference, for example a virtual nose \cite{b5,b10}. Yet in the end it requires human testing to find out how nauseating an application is as soon as any movement is included.
One study \cite{b1} showed selected youtube videos to people in VR and let them rate how nauseating their experience was. The data is used to train a neural network to predict nausea based on mean disparity, mean vertical velocity and medium speed. Since they only have a limited database their data is “obtained from real-world measurements of users”.

\section{The RCVR framework for automatic motion sickness prediction}

The proposed framework consists of three main stages, the rollercoaster track generation and data collection, the input data selection and the machine learning architectures.

\subsection{Creating rollercoasters and collecting in-game data}

\begin{figure}[h!]
		\centering
  \includegraphics[scale=0.3]{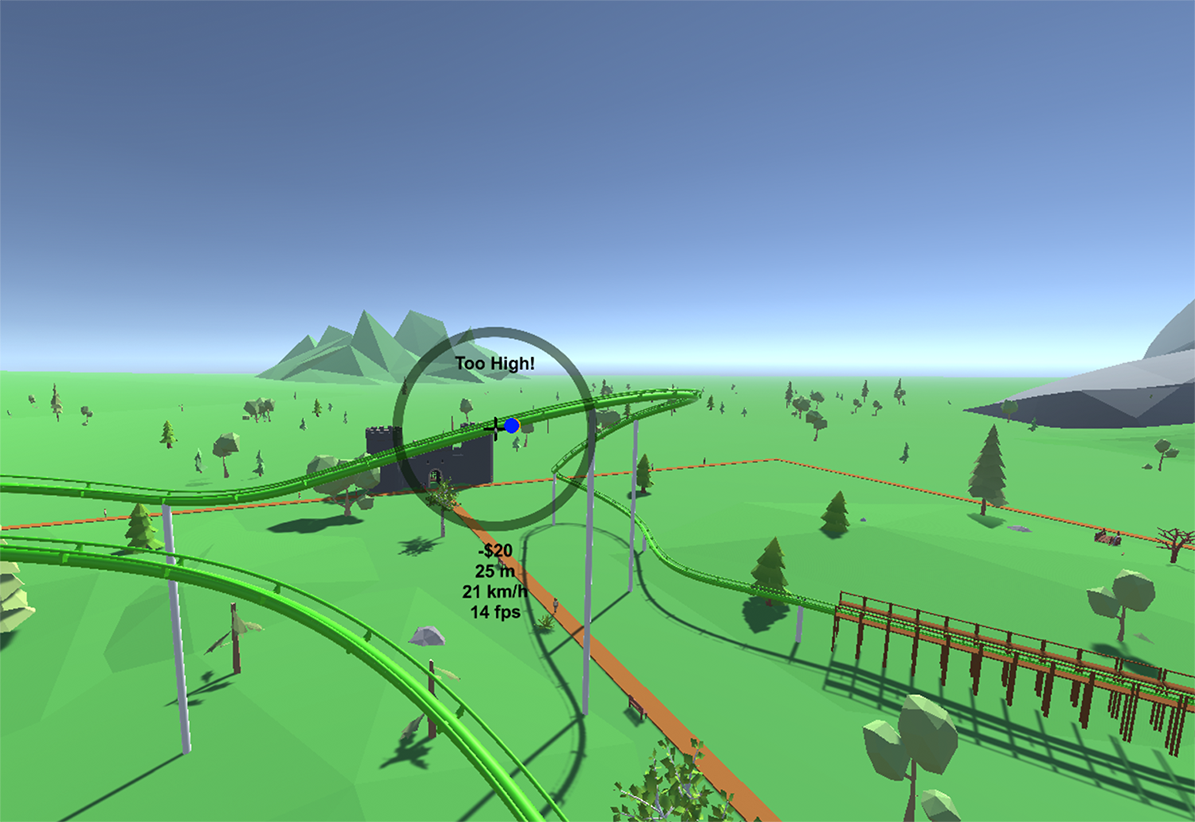}
  \caption{Creating a rollercoaster by tilting your head.}
  \label{fig:steering}
\end{figure}

As head rotation is tracked in the proposed virtual environment users have one more degree of freedom available than when using a mouse. Head rotation can therefore be used well to create a rollercoaster. Similar to a flight simulator the users move in the direction in which they tilt their heads as shown in Figures \ref{fig:rollercoaster} and \ref{fig:steering}. As in a real rollercoaster the players accelerate if they are facing downwards and break if they are facing upwards. The path through the environment creates the rollercoaster track. This provides an intuitive, fast and natural interface to design rollercoasters. The players progress through a number of environments in each of which they have to build a successful theme park. In every park it is required to fulfill certain criteria, like having a rollercoaster with a fun rating of at least $S_F=3$ but a nausea rating of less than $S_N=3$. These different tasks ensure that users create a diverse park with different types of rollercoasters.

\begin{figure}[h!]
		\centering
  \includegraphics[scale=0.3]{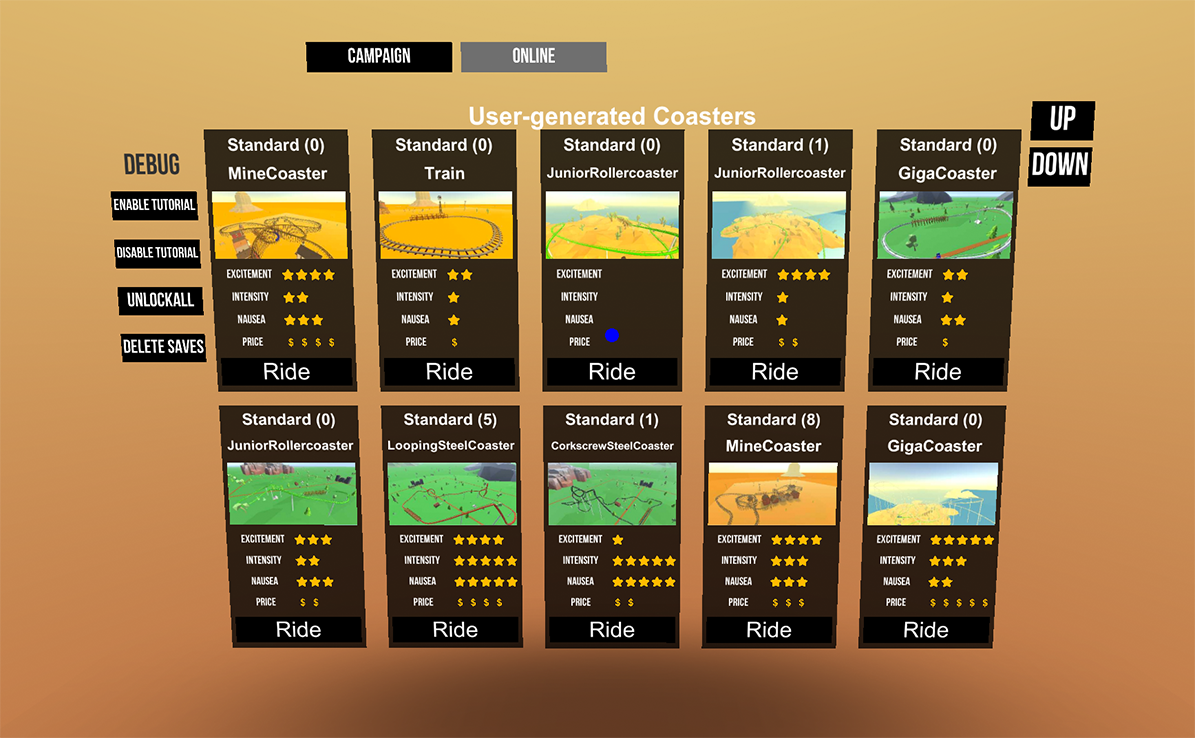}
  \caption{Rollercoaster uploads and user ratings.}
  \label{fig:park}
\end{figure}

The players are encouraged to upload their rollercoasters (see figure \ref{fig:park}) and other users can select any of the uploaded ones and ride them – at the end of the ride they are asked how much fun the ride was, how intense, how nauseating and how much they would pay for this ride in a small Simulator Sickness Questionnaire \cite{b11}. All ratings are given in a range from $1$ to $5$ stars. The ratings and uploaded rollercoasters create the proposed RCVR database containing the users' ratings and the in-game collected data. Rollercoasters are stored as an array of data points which contain the position, rotation, speed and gravitational forces exerted at this point. Besides rollercoaster applications, this can also be applied on racing games, space simulations and many more genres.

\subsection{Preparing the input data and the ground truth}
The core of this project are the input parameters in the proposed neural networks. The different types of networks and input vectors used in this work are analysed below. Several input vector representations of a ride on the rollercoaster were considered. The first most flexible approach is to use a video of the whole rollercoaster ride as input vector for a deep NN architecture. The advantage of using a video is that it is as close to the human experience as possible. Furthermore, the environment is taken into consideration, every object present in the field of view as well the frames of reference (i.e. the shape of the rollercoaster coach). A video-based approach would require a very large database and to infer the motion additional methods are required imposing errors and bias. Since the developed game already provides the motion vectors of the ride this is unnecessary.

The second approach is to feed all stored rollercoaster points (i.e. representation of the virtual environment) into the RCVR dataset. Each point can include:
\begin{itemize}
\item Absolute position.
\item Relative position from previous rollercoaster point
\item Current speed.
\item Gravitational forces exerted at this point.
\end{itemize}

Additionally, from these data, there is a possibility to define custom input parameters that are generally considered to have an important effect on the experience of a rollercoaster ride
\begin{itemize}
\item Maximum speed
\item Average speed.
\item Total length.
\item Maximal downwards angle.
\item Maximal upwards angle.
\item Type of rollercoaster.
\end{itemize}

Gravitational forces are the main contributing factor to a rollercoaster experience. G-forces function on three axis: the vertical axis and the two horizontal axes. For each axis we store the following parameters:

\begin{itemize}
\item Maximal positive force.
\item Maximal negative force.
\item Average positive force (only sampled at points with force greater than zero).
\item Average negative force, Percentage of the ride where a positive force is exerted.
\item Percentage of the ride where a negative force is exerted.
\end{itemize}

Finally, one issue with the proposed dataset is that as different users rate each rollercoaster differently there are multiple data for the same input resulting to different outputs. While this is not necessarily a problem and the neural network will converge towards the average of all ratings for a rollercoaster, it may be more preferable to group the ratings from the questionnaire for each rollercoaster with an algorithm. Several options are available in the neural network trainer of the project. \par

\begin{itemize}
\item Keep all. All ratings are kept and fed into the neural network.
\item Average. All ratings for the same rollercoaster are averaged, the data set is reduced to the number of rollercoasters. The problem with this approach is that the average rating is very rarely at the end of the range (1 star or 5 star), values close to 3 stars are most probable.
\item Choose most-picked. This approach addresses the issue that average ratings tend to stay around 3 stars. The star rating that has been picked most will be taken, all the other values are discarded. For example, if 4 users rate the rollercoaster with 5 stars and 2 users rate it with 1 stars, the result is simply 5 stars.
\end{itemize}

Furthermore, the option to consider computer vision metrics based on global motion estimation \cite{b13} and scene recognition could help to provide more reliable estimates of what triggers motion sickness. In this case deep CNN architectures could be used for the analysis of the effects of VR combined with the scene contents.

\subsection{Network architecture}
\begin{figure}[h!]
		\centering
  \includegraphics[scale=0.95]{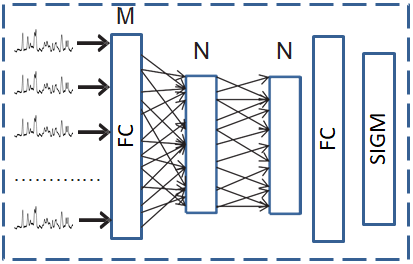}
  \caption{RCVR-net architecture.}
  \label{fig:figNN}
\end{figure}

There are a variety of neural networks available that are useful for different purposes. Convolutional neural networks can analyze more complex problems but typically require a large database or need to be pre-trained. Most pre-trained networks focus on visual tasks like image classification or object recognition and are therefore not suitable to evaluate the time series in the proposed RCVR dataset. Fully connected layers combined with recurrent neural networks and sigmoid ones provide an architecture (see Figure \ref{fig:figNN}) that takes inputs of variable length and is providing the expected nausea level.

\section{Evaluation and Results}
Regarding the training process, 80\% of the data are reserved for training and 20\% for testing using a cross validation approach. At the writing of this paper the database contains 100 ratings from 23 users for 33 rollercoasters. \par

The network contains 20 output neurons - 5 outputs each for fun, intensity, nausea and price levels. Each output neuron represents one out of 5 possible stars - with 5 being the most nauseating experience. This paper focusses on the nausea output of the network. \par

\begin{figure}[h!]
		\centering
  \includegraphics[scale=0.651]{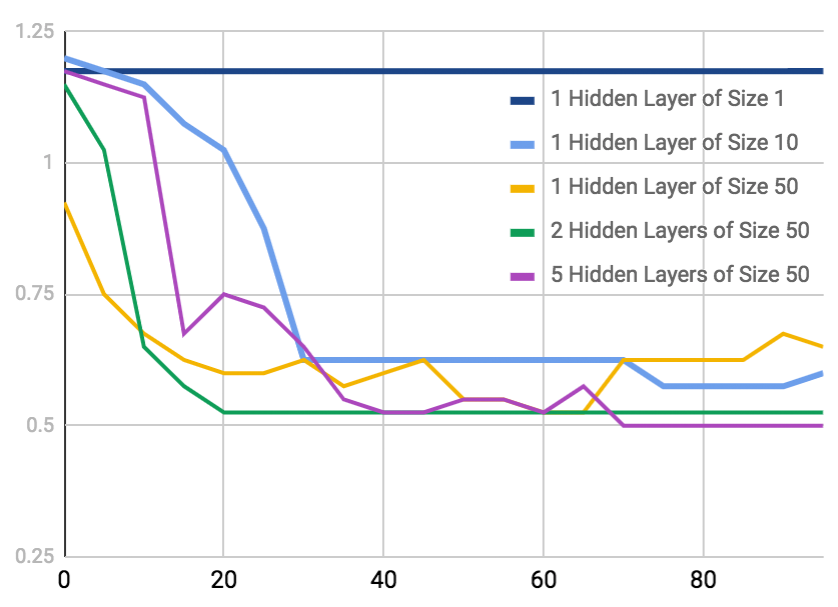}
  \caption{Mean error (y) after x iterations, given different hidden layer sizes and numbers of hidden layers.}
  \label{fig:iterations}
\end{figure}

Table I evaluates different settings for the neural network on the performance of predicting the nausea rating. It shows the percentage of correct answers as well as the mean error (given in stars, with a maximum possible error of 4 stars if the user rating is 1 star and the network rating is 5 stars). Figure \ref{fig:iterations} shows the mean error after every iteration for the proposed neural network. \par

\begin{table}[]
\label{table:results}
\caption{Neural network results for nausea}
\begin{tabular}{@{} p{5cm}|p{1cm}|p{1.5cm}@{}}
 & \%  & mean error  \\
 Twenty five custom input parameters, 20 output parameters, 5 hidden layers of size 40 & 40\%  & 0.75  \\
  Seven input parameters per rollercoaster data piece consisting of relative position, speed and gravitational force, 20 output parameters, 5 hidden layers of size of 50 & 50\%  & 0.6  \\
Combined custom input parameters and rollercoaster data pieces, 5 hidden layers of size 100 & 50\%  & 0.6  \\
\end{tabular}
\end{table}

Using the custom inputs allows the neural network to be much smaller but achieves slightly worse results than feeding in all rollercoaster pieces. Combining both methods does not improve the results (see Table \ref{table:results}). This demonstrates that even with a small dataset it is possible to predict the nausea level to a certain extent. While the neural network struggles to predict how much fun a rollercoaster is the intensity and price value can also be predicted quite well with a neural network.\par
Achieving acceptable results with a small database is reassuring and promising for future extensions. Once the application and the dataset are published it should help other developers and researchers to predict nausea given a path through an environment.

\section{Conclusion}

In this work the problem of predicting motion sickness just from the virtual environment and the expected in-game behaviour was investigated. The proposed neural network architecture provides acceptable results in predicting motion sickness and intensity of rollercoasters. It has been less successful in predicting how fun a rollercoaster is but using a more extended database in the future the neural network will eventually be able to give consistently reasonable ratings for rollercoasters.

The nausea value is the most important value calculated by the network as it evaluates a key challenge of VR applications but the intensity value is also important to hint at if an experience is gentle or fast-paced. The proposed framework and the database could be useful for testing and evaluation purposes in any application or game that involves movement in Virtual worlds. As testers become gradually accustomized to motion sickness this is a field in which machine learning could outperform users in giving more unbiased values and averaging the wide range of human susceptibility to motion sickness.

\vspace{12pt}

\end{document}